\documentclass[preprint]{imsart}
\usepackage{amsthm,amsmath}

\RequirePackage{natbib}
\usepackage{amssymb, amsfonts,mathrsfs,dsfont,bm}   
\usepackage{graphics}
\usepackage{graphicx}
\usepackage{algorithm}
\usepackage{algorithmic}
\usepackage{url}

\usepackage{nicefrac}       

\newtheorem{theorem}{Theorem}

\newtheorem{remark}[theorem]{Remark}
\newtheorem*{theorem*}{Theorem}

\newcommand{\Ecal}{\mathcal{E}}
\newcommand{\Fcal}{\mathcal{F}}
\newcommand{\Xcal}{\mathcal{X}}
\newcommand{\Rb}{\mathbb{R}}
\newcommand{\Nb}{\mathbb{N}}

\newcommand{\AU}{\mathrm{AU}}
\newcommand{\sign}{\mathrm{sign}}

\newcommand{\argmin}{{\rm arg}\mathop{\rm min\,}\limits}

\newcommand{\BP}{\mathrm{BP}}
\newcommand{\SI}{\mathrm{SI}}

\def\ba#1\ea{\begin{align*}#1\end{align*}} 
\def\banum#1\eanum{\begin{align}#1\end{align}} 
\startlocaldefs
\endlocaldefs

\begin{document}
\begin{frontmatter}

\title{Selective inference after feature selection via multiscale bootstrap}
\runtitle{Selective inference of features via multiscale bootstrap}


\author{\fnms{Yoshikazu} \snm{Terada}\ead[label=e1]{terada@sigmath.es.osaka-u.ac.jp}\thanksref{t1}}
\thankstext{t1}{
Jointly affiliated at 
RIKEN Center for Advanced Intelligence Project (AIP),
1-4-1 Nihonbashi, Chuo-ku, Tokyo 103-0027, Japan.
}
\address{
Graduate School of Engineering Science, Osaka University\\
1-3 Machikaneyama-cho, Toyonaka, Osaka 560-8531, Japan\\
\printead{e1}}
\and
\author{\fnms{Hidetoshi} \snm{Shimodaira}\ead[label=e2]{shimo@i.kyoto-u.ac.jp}\thanksref{t1}}
\address{
Graduate School of Informatics, Kyoto University\\
Yoshida Honmachi, Sakyo-ku, Kyoto, 606-8501, Japan\\
\printead{e2}}

\runauthor{Y.~Terada and H.~Shimodaira}

\begin{abstract}
It is common to show the confidence intervals or $p$-values of selected features, or predictor variables in regression, but they often involve selection bias. The selective inference approach solves this bias by conditioning on the selection event. Most existing studies of selective inference consider a specific algorithm, such as Lasso, for feature selection, and thus they have difficulties in handling more complicated algorithms. Moreover, existing studies often consider unnecessarily restrictive events, leading to over-conditioning and lower statistical power. Our novel and widely-applicable resampling method via multiscale bootstrap addresses these issues to compute an approximately unbiased selective $p$-value for the selected features. As a simplification of the proposed method, we also develop a simpler method via the classical bootstrap.
We prove that the $p$-value computed by our multiscale bootstrap method is more accurate than the classical bootstrap method.
Furthermore, numerical experiments demonstrate that our algorithm works well even for more complicated feature selection methods such as non-convex regularization.
\end{abstract}

\end{frontmatter}

\section{Introduction}

In the classical statistical inference, 
the specification of a hypothesis is assumed to be independent of obtained data.
In recent years, since big and complicated data have been common in various fields, 
it is difficult to set hypotheses in advance. 
Thus, in modern data analysis, 
we commonly find useful hypotheses from obtained data using exploratory data analysis,
and then we perform the classical inference for the selected hypotheses.
However, we often ignore the effects of the hypothesis selection in the classical inference, and thus
this naive approach will not provide a valid statistical inference.
Recently, the selective (or post-selection) inference, which deals with the hypothesis selection effect appropriately, 
has drawn considerable attention not only in the statistical community but also in the machine learning community (e.g., \citealp{FanEtAl16, SuzumuraEtAl17, SlimEtAl19, LimEtAl20}).

In this paper, we focus on the selective inference after the feature selection, i.e., predictor variable selection, in regression analysis.
The most intuitive and straightforward approach of selective inference is proposed by \citet{Cox75} and called data splitting.
In data splitting, an i.i.d.\ sample is divided into two subsamples: one is used for the feature selection, 
and the other is used for the inference of the selected features.
However, this approach reduces available data for both feature selection and inference.
\citet{fithian2014optimal} provides the theoretical foundation 
to consider the optimality of the selective inference in the sense of statistical power.
In \citet{BerkEtAl13}, without assuming a specific feature selection method, 
a valid selective inference after feature selection for the submodel parameters is developed on the regression problem.
Importantly, \citet{BerkEtAl13} also introduces both \emph{submodel view} and \emph{full-model view} of the targets of selective inference after feature selection.
Under the setting of \citet{BerkEtAl13}, 
\citet{LeeEtAl16} characterizes the selection event in which a specific model is selected by Lasso (\citealp{Tibshirani96}).
More precisely, this selection event is represented as a union of polyhedra in the space of the response variable.
In addition, based on this fact, \citet{LeeEtAl16} proposes the exact selective inference for the feature selection via Lasso.
The significance levels conditioned on the selection event are computed by truncated normal distributions, justified by the \emph{polyhedral lemma}.
\citet{RyanTibshiraniEtAl16} develops a general framework to perform selective inference after any selection event 
that is represented as a response vector falling into a polyhedral set. 
\citet{RyanTibshiraniEtAl18} proves that this selective inference is asymptotically valid even for non-normal error distributions.

On the other hand, 
the exact selective inference approaches such as \citet{LeeEtAl16} and \citet{RyanTibshiraniEtAl16}
assume that the selection event is explicitly represented as a union of polyhedra in the space of the response variable.
Although the idea of the polyhedral lemma is, in fact, valid for any selective sets beyond a union of polyhedra (\citealp{LiuEtAl18}), 
the existing approaches have computational difficulties in handling more complicated algorithms with non-convex penalties such as MCP (minimax concave penalty; \citealp{Zhang10}) and SCAD (smoothly clipped absolute deviation; \citealp{FanLi01}), where the selective sets become more complicated than the ordinary Lasso.
Although the selective inference of \citet{BerkEtAl13} is not limited to specific feature selection methods, the computation cost may be prohibitive for the number of variables over $20$. 
In addition, it controls type-I errors simultaneously under all submodels, thus leading to very conservative confidence intervals.
Moreover, most existing selective inference with the full-model view is unnecessarily \emph{over-conditioning} and lower statistical power because the inference is conditioned on a selected model, whereas it could be minimally conditioned on a selected feature.
The selective set of the minimally conditioning event becomes more complicated and computationally difficult, and thus its valid post-selection inference is implemented recently by \citet{LiuEtAl18} first time but only for the ordinary Lasso case.

Recently, \citet{TeradaShimodaira17} extends the general hypothesis testing framework, called {\it the problem of regions} (\citealp{EfronTibshirani98}), to the selective inference, 
and proposes a new selective inference approach via {\it multiscale bootstrap} of \citet{Shimodaira:2002:AUT,Shimodaira:2004:AUT,Shimodaira08}.
This approach is not based on the polyhedral lemma, and 
we can easily compute approximately unbiased selective $p$-values for hypotheses conditioned on complicated selective sets.
Moreover, \citet{TeradaShimodaira17} provides the theoretical justification for this approach in two asymptotic theories.
In this framework, we consider the general setting in which
the hypothesis and the selection event are represented as regions in some parameter space.
This approach can be widely applied because we do not need to know the shapes of these regions, but only need to prepare functions that tell whether these regions include a realization of the parameter estimate.
In fact, \citet{ShimodairaTerada19} describes an application of this approach for testing trees and edges in phylogenetics.
Moreover, based on our idea described in this paper and \citet{TeradaShimodaira17}, 
\citet{LimEtAl20} develops the powerful selective inference after feature selection using the Hilbert Schmidt Independence Criterion and the Maximum Mean Discrepancy.

In the original form of multiscale bootstrap method, 
we change the sample size of bootstrap samples 
and then compute a bias-corrected $p$-value using geometric quantities (curvature and signed distance of the region) estimated from the scaling-law of the bootstrap probability of the hypothesis and selection event. 
However, this multiscale bootstrap method cannot be directly applied to selective inference after feature selection 
since the shape of the selective region is unwillingly related to the sample size in the feature selection problem. 
To overcome this difficulty, we propose the use of the resampling of the residuals with scale change. 
The advantage of our method is that it can be applied to almost any feature selection algorithm. 
In addition, the computational complexity of our method is the same order as the classical bootstrap method. 

This paper is organized as follows.
In Section~\ref{sec:SI}, we describe the setting of selective inference after feature selection.
In Section~\ref{sec:mbp}, we give a brief exposition of multiscale bootstrap and 
the general selective inference via multiscale bootstrap.
In Section~\ref{sec:selective_regression},
we develop a new selective inference algorithm via multiscale bootstrap in regression analysis.
In Section~\ref{sec:exp}, 
the usefulness of our approach is demonstrated through numerical experiments.

\section{Selective inference after feature selection}\label{sec:SI}
We briefly describe the selective inference after feature selection in linear regression; we will give the setting in Section~\ref{sec:selective_regression} with details.
Let $Z=(Z_1,\dots,Z_n)^T$ be the response variable with mean $\xi\in\Rb^n$ and variance $\tau^2 I_n$.
Let $x_1,\dots,x_p \in \Rb^n$ be non-random features (i.e., predictor variables) and $X=(x_1,\dots,x_p)=(x_{ij})_{n\times p}$.
At first, we need to clarify the target of statistical inference.
If we assume the following first-order correctness:
\[
\exists \beta^\ast\in \Rb^p;\;\xi = X\beta^\ast,
\]
the target of the estimators is clearly the ``true'' coefficients $\beta^\ast$.
However, in real data analysis, it is difficult to assume this correctness,
as mentioned in Box's famous quote ``All models are wrong''.
Even under the first-order correctness, the selected models may be wrong.
Without the first-order correctness, the target of inference is ambiguous.
\cite{BerkEtAl13} clarifies the target of statistical inference in linear regression, 
and there are two possible types of the target:
\begin{itemize}
\item Let $M\subseteq \{1,\dots, p\}$ be a specified set of features.
We consider a submodel using the features in $M$.
The target for the submodel view with respect to $M$ is
\ba
\beta^{(M)} 
&:=\mathop{\argmin}_{b_M\in \mathbb{R}^{\#(M)}}\mathbb{E}\left[\|Z-X_Mb_M\|^2 \right] 
= (X_M^TX_M)^{-1}X_M^T\xi,
\ea
where $X_M$ is the predictor matrix consisting of the features of $M$.
The true value of $\beta_j^{(M)}\;(j\in M)$ in submodel view depends on $M$.
\item
The target for the full-model view is $\beta = \beta^{(M)}$ with $M=\{1,\ldots,p\}$. Thus
\ba
\beta &=(\beta_1,\dots,\beta_p)
= \mathop{\argmin}_{b\in \mathbb{R}^p}\mathbb{E}\left[\|Z-Xb\|^2 \right]
= (X^TX)^{-1}X^T\xi.
\ea
The true value of $\beta_j$ in the full-model view does not depend on $M$. 
In this paper, the target for our method is the coefficients in the full-model view, while both views are discussed below.
\end{itemize}

Now, we describe the basic concepts of selective inference in regression.
Let $\alpha$ be a significant level in selective inference.
When the null hypothesis $H_0$ is selected based on data, 
we should control the selective type I error rate:
\banum
P_{H_0}(H_0\text{ is rejected}\mid H_0\text{ is selected})\le \alpha,\label{eq:selective-type-I-error}
\eanum
where $P_{H_0}$ is a probability distribution under $H_0$.
The event $\{H_0\text{ is selected}\}$ is called the selection event.
After feature selection based on data,
we obtain the selected model $\hat{M}\subseteq\{1,\dots,p\}$.
Then, depending on whether the target is in the submodel view or the full-model view, the null hypotheses for $j \in \hat{M}$ is
$H_0: \beta_j^{(\hat M)}=0$ or $H_0: \beta_j = 0$, respectively.
Here $\beta_j^{(\hat M)}=0$ is a simplified notation of $ \beta_j^{(M)}=0$ with $\hat M = M$.
Moreover,
we may consider two different types of events 
$\{j \in \hat{M}\}$ or $\{\hat{M}=M\}$ 
as the selection event.

However, the event $\{j \in \hat{M}\}$ is not an appropriate selection event
for the hypothesis of submodel view $H_0: \beta_j^{(\hat M)} = 0$, because this hypothesis depends on the other selected features $\hat M \setminus \{j\}$ and thus the probability (\ref{eq:selective-type-I-error}) does not make sense.
Therefore, for the hypothesis of submodel view, the event $\{\hat{M}=M\}$ or more restrictive events are appropriate as a selection event;
the event $\{\hat{M}=M, \mathrm{sign}(\hat{\beta}^{(M)}) = s\}$ is sometimes considered as the selection event (e.g., \citealp{LeeEtAl16}) 
for computational reason.

On the other hand, 
we can consider the two different types of conditioning $\{j \in \hat{M}\}$ and $\{\hat{M}=M\}\;(j \in M)$ as a selection event
for the hypothesis of full-model view $H_0: \beta_j = 0$.
Since both of these two events are appropriate, we may wonder which of them is more desirable.
This is answered by the argument of the monotonicity of selective error in Proposition~3 of \cite{fithian2014optimal}. Here we see its adaptation to our setting.
Since $\{j \in \hat M\} = \bigcup_{M: j\in M} \{\hat M = M\}$, we have
\ba
&P_{H_0}(H_0: \beta_j = 0\text{ is rejected}\mid j \in \hat{M})\\
&=
\sum_{M: j \in M}
P_{H_0}(H_0: \beta_j = 0\text{ is rejected}\mid \hat{M}=M)
\frac{P_{H_0}(\hat{M} = M)}{P_{H_0}(j \in \hat{M})}\\
&\le
\max_{M: j \in M}
P_{H_0}(H_0: \beta_j = 0\text{ is rejected}\mid \hat{M}=M).
\ea
If we control the selective type-I error $P_{H_0}(H_0: \beta_j = 0\text{ is rejected}\mid \hat{M}=M)$ at level $\alpha$ for all models $M\;(j\in M)$, 
the selective type-I error $P_{H_0}(H_0: \beta_j = 0\text{ is rejected}\mid j \in \hat{M})$ is also automatically controlled at level $\alpha$.
Thus, this monotonicity tells us that the over-conditioning leads to a loss of information and that, for the hypothesis  $H_0: \beta_j = 0$,  
the minimal selection event $\{j \in \hat{M}\}$ (i.e., the minimally conditioning and thus the maximal event set) is the most desirable in the sense of statistical power. 

\section{An overview of multiscale bootstrap}
\label{sec:mbp}

\begin{figure}
  \centerline{ \includegraphics[width=0.98\textwidth]{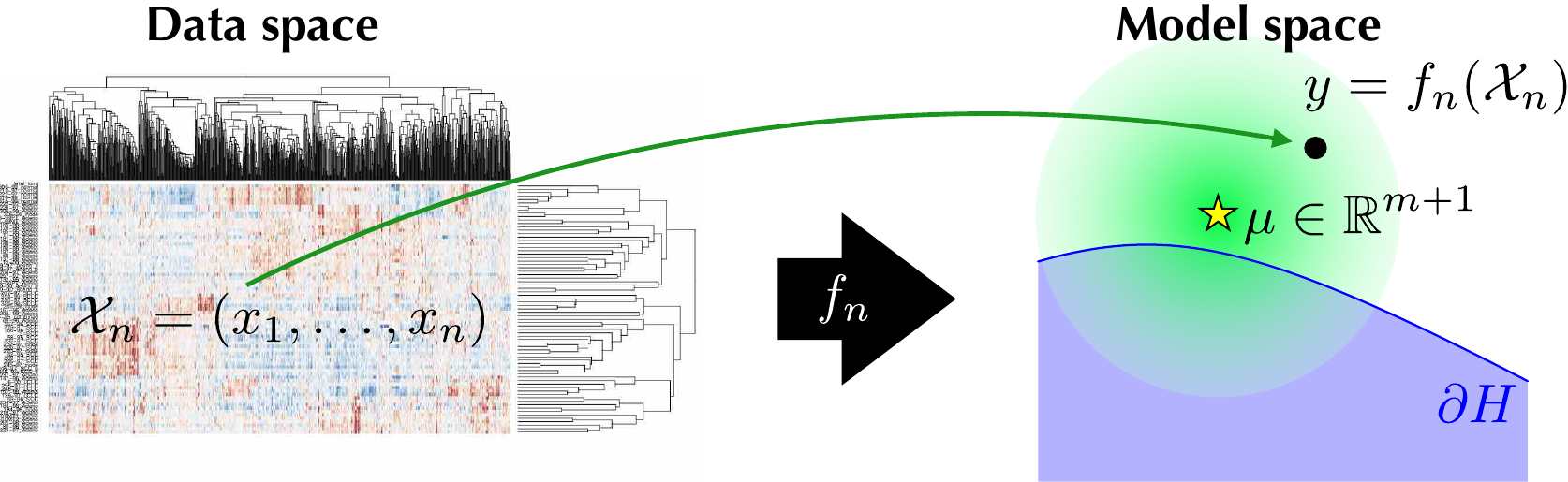}}
  \caption{Relationship between the data space and the model space.}
  \label{fig:model space}
\end{figure}

First, we describe the framework of the problem of regions in Section~\ref{sec:probllem-of-regions}, followed by the basic idea of multiscale bootstrap for non-selective inference
(\citealp{Shimodaira:2002:AUT,Shimodaira:2004:AUT,Shimodaira08}) in Section~\ref{sec:ordinary-multiscale-bootstrap}.
Then we briefly introduce the general selective inference framework proposed by \citet{TeradaShimodaira17} in Section~\ref{sec:selective-multiscale-bootstrap}.

\subsection{The problem of regions} \label{sec:probllem-of-regions}
The general statistical inference framework, in which the hypothesis is represented by a general region in some parameter space,
 is called {\it the problem of regions} (\citealp{EfronTibshirani98}).
 This framework is an abstraction of many applications, e.g., phylogenetic inference, in which a confidence level is assigned for each clade of  the estimated phylogenetic tree (\citealp{Felsenstein:1985:CLP,Efron:Halloran:Holmes:1996:BCL}).

Let $\Xcal_n = (X_1,\dots,X_n)$ be a data with sample size $n$.
In the problem of regions, it is assumed that 
there exists a transform $f_n$ of $\Xcal_n$ such that the transformed data follows the $(m+1)$-dimensional Gaussian distribution with unknown mean parameter $\mu\in\mathbb{R}^{m+1}$ and covariance identity $I_{m+1}$:
\[
Y:=f_n(\Xcal_n) \sim N_{m+1}(\mu, I_{m+1}).
\]
Typically, $f_n$ involves multiplying the factor $\sqrt{n}$ to a form of sample average so that the covariance matrix of $Y$ is properly rescaled. 
Here, the $(m+1)$-dimensional space of $Y$ will be referred to as the model space in this paper.
Figure~\ref{fig:model space} shows the image of the model space.
In addition, let $y\in\mathbb{R}^{m+1}$ be an observed value of $Y$, and suppose that 
a bootstrap sample $\mathcal{X}_{n^\prime}^\ast=(X_1^\ast,\dots,X_{n^\prime}^\ast)$ with sample size $n^\prime$ is represented as a realization of the following Gaussian distribution
in the model space:
\[
Y^\ast = f_{n}(\Xcal_{n^\prime}^\ast)\sim N_{m+1}(y,\sigma^2I_{m+1}),\quad\sigma^2 = \frac{n}{n^\prime}.
\]
We will denote by $P_{\sigma^2}(\cdot|y)$ the probability measure of the bootstrap sample $Y^\ast$ with scale $\sigma>0$.
This framework is a simplification of reality and is justified by the central limit theorem in many situations.

Let $H \subset \mathbb{R}^{m+1}$ be a general region and let us consider $H_0:\mu \in H$ as a null hypothesis.
It is assumed that the region $H$, in a neighbourhood of the model space, is locally represented as $H=\{(u,v)\mid v \le -h(u),\; u \in \Rb^{m}, v \in \Rb\}$ using some continuous function $h:\Rb^m \rightarrow \Rb$.
Let $\partial H := \{(u,v)\mid v = -h (u),\; u \in \Rb^{m}\}$ be the boundary surface of the region $H$.
In this setting, our main goal is to compute an approximately unbiased $p$-value $p(H|y)$ for the null hypothesis $H_0:\mu \in H$ against the alternative hypothesis $H_1:\mu \not\in H$. The approximately unbiased $p$-value should satisfy
\begin{align}
\forall \mu \in \partial H;\;P(p(H|Y)<\alpha\mid \mu) \approx \alpha\label{eq:au}
\end{align}
for a given significance level $\alpha>0$.
In other words, the $p$-value is approximately distributed as uniform over (0,1), i.e., $p(H|Y) \sim U(0,1)$ when $\mu\in \partial H$.
The difference between $P(p(H|Y)<\alpha\mid \mu)$ and $\alpha$ in (\ref{eq:au}) is called {\it bias} (or error).
The bootstrap probability
\[p_{\BP}(H|y):=P_{1}(Y^\ast \in H|y)
\]
is considered as the most simple $p$-value satisfying $(\ref{eq:au})$; see \citet{Efron:Halloran:Holmes:1996:BCL,EfronTibshirani98}.
More formally, in the classical large sample theory, if the region $H$ has a smooth boundary surface, 
the bootstrap probability $p_{\BP}(H|y)$ has the first-order accuracy: 
$
\forall \mu \in \partial H;\;P(p_{\BP}(H|Y)<\alpha\mid \mu) = \alpha + O(n^{-1/2}).
$
However, in many practical situations, the bootstrap probability $p_{\BP}$ often has a non-negligible bias.
 
 \subsection{Basic idea of multiscale bootstrap}
\label{sec:ordinary-multiscale-bootstrap}

To obtain more accurate $p$-values, 
geometric quantities in the model space, such as distance and curvature, play a key role. 
In fact, \citet{EfronTibshirani98} shows that we can compute a more accurate $p$-value using the signed distance $v(H|y)$ from the data point $y$ to the region $H$.
More precisely, the $p$-value $p_{\sign}(H|y) := P_1(v(H|Y^\ast) \ge v(H|y)\mid \hat{\mu}(y))$ is proposed,
where $\hat{\mu}(y)\in\partial H$ is the projection point of $y$ onto $\partial H$.
This $p$-value $p_{\sign}(H|y)$ has the third-order accuracy (\citealp{Efron:1985:BCI,EfronTibshirani98}).

However, in most practical situations, 
it is difficult to access the model space and to obtain the explicit formula of the hypothesis region in the model space.
Thus, we cannot compute the signed distance $v(H|y)$ in general.
To overcome this difficulty, \citet{Shimodaira:2002:AUT,Shimodaira:2004:AUT,Shimodaira08} propose a new bootstrap method, called {\it multiscale bootstrap}.
In multiscale bootstrap, 
the geometric quantities such as the signed distance $v(H|y)$ and the mean curvature of $\partial H$ are estimated based on the scaling law of the bootstrap probabilities, 
and an accurate $p$-value is computed based on these estimated quantities.

We consider the bootstrap probability with scale $\sigma>0$
\[
\alpha_{\sigma^2}(H|y) := P_{\sigma^2}(Y^\ast \in H | y),
\]
which reduces to $p_{\BP}(H|y)$ when $\sigma=1$.
When we change $n'$ in the data space, and in effect, change $\sigma$ in the model space, the bootstrap probability changes.
This change is simply expressed in terms of the normalized bootstrap $z$-value defined as
\[
\psi_{\sigma^2}(H|y):=\sigma \bar{\Phi}^{-1}(\alpha_{\sigma^2}(H|y)),
\]
where $\Phi(x)$ is the cumulative distribution function of the standard normal distribution and $\bar{\Phi}(x)=1-\Phi(x)$, i.e., $\bar \Phi^{-1}(\alpha)$ is the upper $\alpha$-value.
\citet{Shimodaira:2002:AUT,Shimodaira:2004:AUT} show
the following scaling law of the bootstrap probabilities:
\[
\psi_{\sigma^2}(H|y) = v(H|y) + \gamma(H|y)\sigma^2 + O_p(n^{-1}),
\]
where $\gamma(H|y)$ is the mean curvature of the boundary surface $\partial H$ at $\hat{\mu}$.
This scaling law can be modelled as the simple linear regression $\theta_{H,0} + \theta_{H,1}\sigma^2$ with $\sigma^2$ as the predictor.
 We will denote by $\varphi_H(\sigma^2|\theta_H)$ the model for the normalized bootstrap $z$-value, such as $\theta_{H,0} + \theta_{H,1}\sigma^2$ with parameter $\theta_H=(\theta_{H,0}, \theta_{H,1})$.
The bootstrap probabilities with several values $\{\sigma_j\}$ of scale can be computed by using the bootstrap samples with different sample sizes, say $n^\prime_j \in \{ \lceil 0.5n \rceil, \cdots , \lceil 1.0n \rceil, \cdots,\lceil 1.5n \rceil \}$.
Let $B$ be the number of bootstrap replicates, and $C_H=\#\{Y^\ast \in H\}$ be the frequency to be $Y^\ast \in H$.
Let $\hat{\psi}_{\sigma_j^2}(H|y)$ be the estimated normalized bootstrap $z$-value by using the estimated bootstrap probability $\hat{\alpha}_{\sigma^2}(H|y) = C_H/B$.
We can estimate the values of $v(H|y)=\theta_{H,0}$ and $\gamma(H|y)=\theta_{H,1}$ by the simple regression for the observed $\{(\sigma_j^2, \hat{\psi}_{\sigma_j^2}(H|y))\}$.

\citet{Shimodaira:2002:AUT} proposes the following $p$-value:
\ba
p_{\AU}(H|y) 
&:= \bar{\Phi}(\varphi_H(-1|\theta_H)) 
= \bar{\Phi}(v(H|y) - \gamma(H|y) ) + O_p(n^{-1}).
\ea
This $p$-value $p_{\AU}(H|y)$ has the second-order accuracy (\citealp{Shimodaira:2004:AUT, EfronTibshirani98}):
\[
\forall \mu \in \partial H;\;P(p_{\AU}(H|Y)<\alpha\mid \mu) = \alpha + O(n^{-1}).
\]
It becomes third-order accurate erring only $O(n^{-3/2})$ when $\varphi_H(\sigma^2|\theta_H)$ is properly estimated from observed values of $\psi_{\sigma^2}(H|y)$ including terms of order $O_p(n^{-1})$.

In the classical large sample theory, 
the shape of $H$ in the model space is magnified by $\sqrt{n}$, and thus
the key property is that the smooth boundary surface $\partial H$ approaches a flat surface 
in a neighborhood of any point on $\partial H$.
In contrast, for non-smooth surfaces, this key property is not satisfied.
For example, if the region $H$ is cone-shaped, 
the shape of $H$ is scale-invariant in a neighborhood of the vertex of $H$.
It is well known that there is no unbiased test for a hypothesis region with a non-smooth boundary (\citealp{Lehmann:1952:TMH}).
To deal with general regions with non-smooth boundary surfaces, 
 \citet{Shimodaira08} develops a new theoretical framework, called the asymptotic theory of {\it nearly flat surfaces}.
In this framework, we require that the magnitude of boundary surfaces is small.
Thus, this framework works well even for non-smooth boundary surfaces when the magnitude of boundary surfaces is not so large, at least locally.
We can interpret that the given surface is on the way to approaching the flat surface in this framework.
This idea is similar to one behind the local alternative framework (\citealp{Lehmann:1999:ELST}), but the rescaling is applied only to the direction normal to the boundary surface while the scale is fixed for the other directions.
 A brief introduction of this theory is provided in Appendix~\ref{sec:A}.

\subsection{General selective inference via multiscale bootstrap} \label{sec:selective-multiscale-bootstrap}

Here, we describe an extended framework of the problem of regions for the selective inference.
In the model space, two regions $H=\{(u,v)\mid v\le -h(u),\;u\in \Rb^{m}, v \in \Rb\}$ and $S=\{(u,v)\mid v> -s(u),\;u\in \Rb^{m}, v \in \Rb\}$ are considered.
Suppose that the selection event is represented as $\{ y\in S \}$, 
and we consider the selective inference in which the null hypothesis $H_0:\mu\in H$ is selected if and only if $y\in S$.
In this setting, for a given significance level $\alpha$, we want to compute selective $p$-values $p(H|S,y)$ satisfying
\[
\forall \mu \in \partial H;\;\frac{P(p(H|S,Y)<\alpha\mid \mu)}{P(Y\in S\mid \mu)} \approx \alpha.
\]

In other words, $p(H|S,Y) \sim U(0,1)$  conditioned on $Y\in S$ when $\mu \in \partial H$.
\citet{TeradaShimodaira17} proposes the following approximately unbiased selective $p$-value $p_{\SI}(H|S,y)$ for regions $H$ and $S$ with smooth boundary surfaces:
\[
p_{\SI}(H|S,y) := \frac{\bar{\Phi}(\varphi_H(-1|\theta_{H}))}{\bar{\Phi}(\varphi_H(-1|\theta_{H}) + \varphi_S(0|\theta_{S}))},
\]
where $\varphi_S$ is the model for the normalized bootstrap $z$-value related to the selective region $S$, 
and $\theta_S$ is the parameter of the model $\varphi_S$.

\begin{theorem} (Theorem 4.3 in \citet{TeradaShimodaira17})
The boundary surfaces $\partial H$ and $\partial S$ are assumed to be sufficiently smooth and nearly parallel in the sense that the first derivatives of $h$ and $s$ differ only $O(n^{-1})$.
Then, the selective $p$-value $p_{\SI}(H|S,y)$ has the second-order accuracy: 
\[
\forall \mu \in \partial H;\;\frac{P(p_{\SI}(H|S,Y)<\alpha\mid \mu)}{P(Y\in S\mid \mu)} = \alpha + O(n^{-1}).
\]
\end{theorem}
The detailed calculation of $p_{\SI}(H|S,y)$ is provided as Algorithm~\ref{alg:simbp}. 


In \citet{TeradaShimodaira17}, 
the selective $p$-value for the regions with non-smooth boundary surfaces is also proposed, 
and the theoretical justification of this $p$-value is provided using the asymptotic theory of nearly flat surfaces.
For more details about the case in which the regions $H$ and $S$ have possibly non-smooth boundary surfaces, 
see Appendix~\ref{sec:A}.
Since the boundary surface $\partial S$ of the selection region is generally not smooth in the selective inference after feature selection, 
the theory of nearly flat surfaces is used to derive the properties of the proposed method described in Section~\ref{sec:si}.
Roughly speaking, both theories essentially assume that
the boundary surfaces $\partial H$ and $\partial S$ are more or less flat and parallel to each other, at least locally around the data point.
In analogy with the local alternative framework, we consider that
given surfaces are on the way to approaching mutually parallel flat surfaces.

\section{Selective inference after feature selection via multiscale bootstrap} \label{sec:selective_regression}

\subsection{Model space for regression analysis}
\label{subsec:model-space}
Here, we describe the setting of the selective inference after feature selection in regression analysis.
We employ the general assumption used in \citet{BerkEtAl13}, \citet{LeeEtAl16}, and \citet{RyanTibshiraniEtAl16}.
Consider the response variable $Z = (Z_1,\dots,Z_n)$ drawn from the multivariate Gaussian distribution:
\[
Z \sim N_n(\xi,\tau^2I_n),
\]
where $\xi \in \Rb^n$ is an unknown parameter, $I_n$ is the $n$-dimensional identity matrix, 
and $\tau^2$ is assumed to be known.
We will denote by $z\in \Rb^n$ the observed value of $Z$.
Let $X=(x_1,\dots,x_p)=(x_{ij})_{n\times p}$ be a non-random full rank matrix whose columns represent the features.
Note that the error variance $\tau^2$ can be estimated if $\xi$ is modeled as a function of features $x_1,\dots,x_p \in \Rb^n$.
Assuming a specific feature selection method, such as Lasso and MCP, is applied to $(X,z)\in \Rb^{n\times p}\times \Rb^n$, let $\hat{M}\subseteq \{1,\dots,p\}$ be the set of selected features, and
$\hat{s}_j \in \{+,-\}$ be the sign of the estimated coefficient $\hat{\beta}_j$ of the feature $j \in \hat M$.

First, note that the general selective inference approach described as Algorithm~\ref{alg:simbp} cannot be directly applied to regression analysis.
Since we need to change the sample size $n'$ of bootstrap samples in the usual multiscale bootstrap, 
it is assumed that the hypothesis and selective regions can be represented as specific regions, which are independent of $n'$, 
in the model space.
For the selective inference in regression analysis, however, 
the shape of the selective region inevitably depends on $n'$ because it is the dimensionality of the model space as explained below.

We recall that 
it is assumed that $Z\sim N_n(\xi,\tau^2I_n)$ and that the selection event can be represented as the region of the space of $Z$.
Then, it is realized that the normalized space of 
\begin{align} \label{eq:model-space-n}
Y := Z/\tau,\quad \mu := \xi/\tau
\end{align}
can be considered as the model space described in Section~\ref{sec:mbp} with $m+1 = n$. Thus, the selective region for multiscale bootstrap inevitably depends on $n'$.
Another choice of model space is given by
\[
 Y:=\tau^{-1} B Z,\quad  \mu:=\tau^{-1} B \xi,
\]
where $B = (X^T X)^{-1/2} X^T \in \mathbb{R}^{p\times n}$.
The selective region represented in this model space also depends on $n'$ because feature selection algorithms take account of sample size.
Although the latter model space is preferable for the asymptotic theory because it has the fixed dimensionality $m+1 = p$, 
we use the former model (\ref{eq:model-space-n}) below for easy illustration.

\subsection{Appropriate selection event}
\label{sec:pre}

\begin{algorithm}[t]      
\caption{Computing approximately unbiased $p$-values for general regions $H$ and $S$}         
\label{alg:simbp}
\begin{algorithmic}[1]
\STATE Specify several $n^\prime\in \Nb$ values, and set $\sigma^2=n/n'$ for each $n'$. 
Set the number of bootstrap replicates $B$, say, 10000.
\STATE For each $n'$, perform bootstrap resampling to generate $Y^*$  for $B$ times and compute $\alpha_{\sigma^2}(H | y)=C_H/B$ and $\alpha_{\sigma^2}(S | y)=C_S/B$ by counting the frequencies
$C_H = \#\{Y^\ast\in H\}$ and $C_S = \#\{Y^\ast\in S\}$.
We may actually work on $\Xcal_{n'}^\ast$ instead of $Y^\ast$.
Compute $\psi_{\sigma^2}(H|y) = \sigma\bar\Phi^{-1} ( \alpha_{\sigma^2}(H|y))$ and
$\psi_{\sigma^2}(S|y) = \sigma\bar\Phi^{-1} ( \alpha_{\sigma^2}(S|y))$.
\STATE Estimate parameters $\theta_H(y)$ and $\theta_S(y)$ by fitting models
$\psi_{\sigma^2}(H | y)=\varphi_{H}(\sigma^2 | \theta_H)$ and 
$\psi_{\sigma^2}(S | y)=\varphi_{S}(\sigma^2 | \theta_S)$,
respectively.
The parameter estimates are denoted as $\hat{\theta}_H(y)$ and $\hat{\theta}_S(y)$.
If we have several candidate models, apply above to each and choose the best model based on AIC value.
\STATE Approximately unbiased $p$-values of non-selective inference ($p_\mathrm{AU}$) and of
selective inference ($p_\mathrm{SI}$) are computed by
$p_{\mathrm{AU}}(H|y) = \bar{\Phi}(z_H )$ and 
$p_{\mathrm{SI}}(H | S, y) = \bar{\Phi}(z_H )/\bar{\Phi}( z_H  + z_S  )$,
where $z_H = \varphi_{H}(-1 | \hat{\theta}_H(y))$ and $z_S =  \varphi_{S}(0 | \hat{\theta}_S(y))$.
\end{algorithmic}
\end{algorithm}

Recently, \citet{LiuEtAl18} suggests the use of the selection event $\{j \in \hat{M}\}$ for a specified $j\in \{1,\ldots,p\}$, which increases the statistical power and thus leading to shorter confidence intervals.
This is explained by the monotonicity of the selective error provided in \citet{fithian2014optimal}, as mentioned in Section~\ref{sec:SI}.
The event $\{\hat M = M\}$ for a specified $M\subset\{1,\ldots,p\}$ is over-conditioning and reducing the statistical power 
because the other features $M \setminus \{j\}$ are not relevant
for the null hypothesis $H_0: \beta_j=0$ of the feature $j$ in the full-model view.

Here, we actually consider testing feature $j$ with its sign.
More precisely, whenever the feature $j$ is selected and $\hat{\beta}_j > 0$ (or $<0$), the hypothesis $H_0:\beta_j \le 0$ (or $\ge0$) is tested.
The minimal selection event is then $\{j\in \hat{M},\hat{s}_j=s_j\}$ where $s_j\in\{+, -\}$.
Hence, the main goal of our selective inference is to compute the unbiased selective $p$-value $p_j(y)$, 
which satisfies 
\banum
\mathbb{P}( p_j(Y) <\alpha \mid j\in \hat{M}\text{ and } \hat{s}_j = s_j) = \alpha \label{eq:unbiased}
\eanum
for any $\mu$ with $\beta_j = 0$.


\subsection{Computing selective $p$-values by multiscale bootstrap}
\label{sec:si}

In this section, we describe our proposed method.
We develop a new algorithm to compute the approximately unbiased selective $p$-value, which approximately satisfies the equation $(\ref{eq:unbiased})$.
We will update the computation of $\psi_{\sigma^2}(H|y)$ and $\psi_{\sigma^2}(S|y)$ in Algorithm~\ref{alg:simbp} to obtain Algorithm~\ref{alg:main}.

\begin{figure*}
\centerline{ 
 \begin{minipage}[t]{0.49\textwidth}
	\centering
  	\includegraphics[width=.98\textwidth]{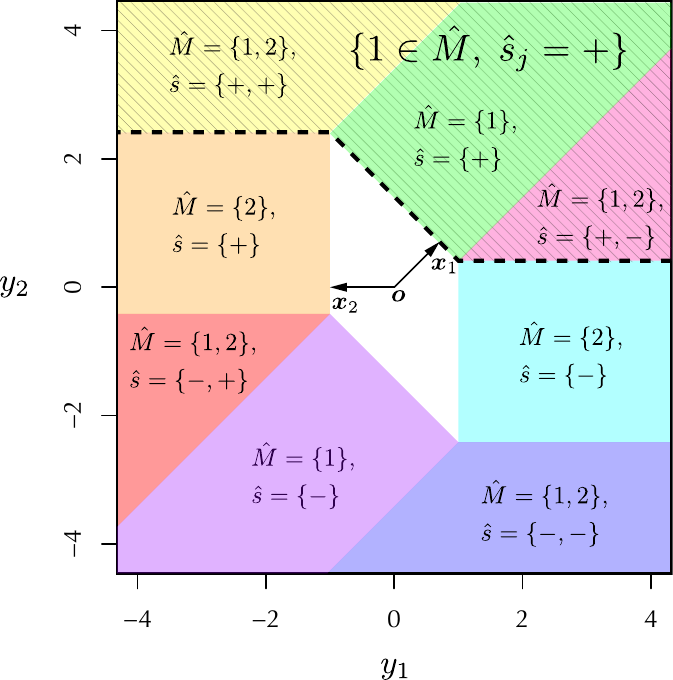}
	(a)~Lasso
  \end{minipage}
 \begin{minipage}[t]{0.49\textwidth}
	\centering
 	 \includegraphics[width=.98\textwidth]{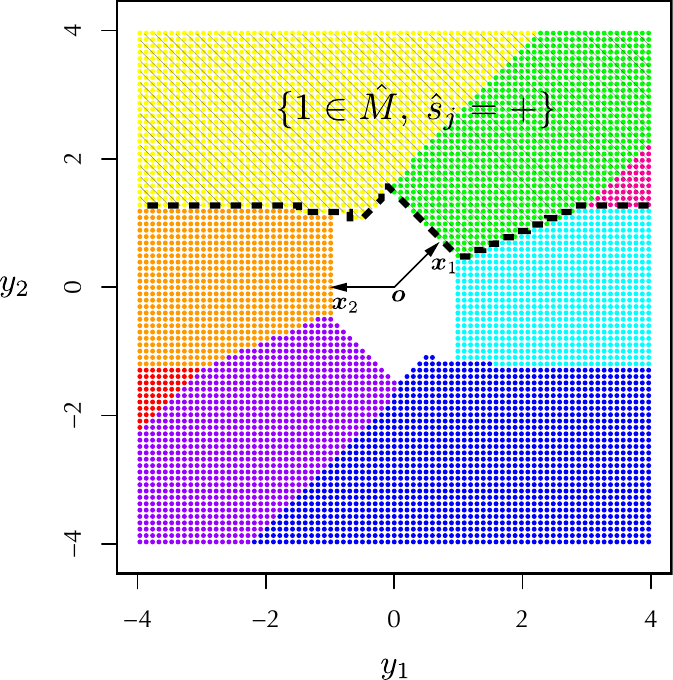}
	(b)~MCP (numerically evaluated at each point)
  \end{minipage}
  }
  \caption{Selective regions of feature selection ($n=2$). The shaded region is corresponding to the event $\{1\in \hat{M},\; \hat{s}_1 = +\}$) which is a union of regions for the selection events of the form $\{\hat M = M, \hat s_M = s_M \}$.
  (a)~For Lasso, each region is represented as a polyhedron.
  (b)~For MCP, each region has a more complicated shape.
  Our selective inference method works easily in both cases.}
  \label{fig:lasso}
\end{figure*}

For the feature selection via Lasso, 
the selection event $\{j \in \hat{M},\hat{s}_j = s_j\}\;(j\in \{1,\ldots,p\},\,s_j\in \{+, -\})$ can be represented as a union of polyhedra 
in the $n$-dimensional space of the response variable (\citealp{LeeEtAl16}).
The left panel of Figure~\ref{fig:lasso} shows the relationship between the selected model by Lasso and the corresponding region in the response vector space when $n=2$.
In contrast, for more complicated feature selection methods such as MCP and SCAD, 
the region $S$ of the selective event $\{j \in \hat{M},\hat{s}_j = s_j\}$ will become complicated, 
and the explicit shape of the selective region $S$ may not be obtained easily.
The right panel of Figure~\ref{fig:lasso} shows the relationship between the selected model by MCP and the corresponding region in the response vector space when $n=2$.
We had to numerically evaluate which features are selected at each point since no explicit representation of the selection event is available.
Although \citet{LeeEtAl16} and \citet{RyanTibshiraniEtAl16} consider exact selective inference for Lasso, it is difficult to consider exact selective inference for these complicated feature selection methods.

%


As shown in Figure~\ref{fig:lasso}, 
the selective region $S$ which represents the selection event $\{j \in \hat{M},\hat{s}_j = s_j\}$ 
could be complicated and has generally non-smooth boundary surfaces.
In contrast, for $\eta\in \Rb$, the hypotheses $H_0:\beta_j \lesseqgtr \eta$, namely the two cases of  $\beta_j \le \eta$ and  $\beta_j \ge \eta$, can be represented as the following regions in the space of $Y$:
\ba
H &= \{z/\tau \in \Rb^n\mid a_j^Tz \lesseqgtr \eta\},\quad
A = 
[a_1, \dots, a_p]^T
=(X^TX)^{-1}X^T.
\ea
Since the hypothesis region $H$ has a flat boundary surface
with mean curvature $\gamma(H|y)=0$, we can easily obtain the expression of $\psi_{\sigma^2}(H|y)$ without multiscale bootstrap.
In particular, for $H_0:\beta_j \lesseqgtr 0$, 
\begin{align} \label{eq:psi-h-flat}
 \psi_{\sigma^2}(H|y) = v(H|y)= \pm \tau^{-1} a_j^Tz / \|a_j\|_2,   
\end{align}
where $v(H|y)$ is the signed distance from $y$ to the hypothesis region $H$.
Thus, once we obtain the expression of $\psi_{\sigma^2}(S|y)$, 
we can compute the selective $p$-value $p_\mathrm{SI}(H|S,y)$.

Next, we consider how to obtain the expression of $\psi_{\sigma^2}(S|y)$ for the selective region $S$ by multiscale bootstrap; 
the bootstrap probability of $S$ is computed at several scales 
via an adaptation of Step 2 of Algorithm~\ref{alg:simbp}.
In the usual setting of multiscale bootstrap, 
changing $n'$ in the data space corresponds to changing $\sigma$ in the model space.
Thus, based on bootstrap probabilities with several $n'$,
we can estimate the expression of $\psi_{\sigma^2}(S|y)$.
However, as described in Section~\ref{subsec:model-space},
this framework is not applicable here.
Fortunately, since we can access the model space, 
we directly change the scale $\sigma$ in the model space 
and compute the bootstrap probabilities with several scales.
With the normality of response $Z$, 
the parametric bootstrap method, i.e., sampling directly from $N_n(y,\sigma^2 I_n)$, could be applied to the computation of 
bootstrap probabilities $\alpha_{\sigma^2}(S|y)$ at several scales $\sigma>0$.
To relax the Gaussian assumption, here we consider the resampling of  residuals with scale change.
More formally, we resample the scaled residuals to compute $\alpha_{\sigma^2}(S|y)$ at several $\sigma>0$ as follows.
Let $\hat{\beta}^{(\mathrm{LS})}=(X^TX)^{-1}X^Tz$ be the least-squares estimator based on the full-model.
Write $\hat{e} := z - X\hat{\beta}^{(\mathrm{LS})}$ and $(h_1,\dots,h_n):=\mathrm{diag}(X(X^TX)^{-1}X^T)$.
Then, the adjusted residuals $\hat{\epsilon} = (\hat{\epsilon}_1,\dots,\hat{\epsilon}_n)^T$ are defined as 
$\hat{\epsilon}_i = \hat{e}_i/\sqrt{1-h_{i}}$.
To compute the bootstrap probability $\alpha_{\sigma^2}(S|y)$ at $\sigma>0$, 
we use the following bootstrap sample:
\banum\label{eq:resdual}
z_\sigma^\ast = X\hat{\beta}^{(\mathrm{LS})} + \sigma \hat{\epsilon}^\ast,
\eanum
where $\hat{\epsilon}^\ast = (\hat{\epsilon}_1^\ast,\dots,\hat{\epsilon}_n^\ast)^T$ is a bootstrap sample with size $n$ from $(\hat{\epsilon}_1,\dots,\hat{\epsilon}_n)$.
For each $\sigma>0$, we generate $z_\sigma^\ast$ for $B$ times, 
and apply a particular model selection procedure to them for computing $\alpha_{\sigma^2}(S|y) = C_S/B$ by counting the frequency of the selective event $\{j \in \hat M, \hat s_j = s_j\}$.

\begin{algorithm}[h]              
\caption{Approximately unbiased selective $p$-value for the selected feature $j \in \hat{M}$ with $\hat{s}=s_j$ }         
\label{alg:main}
\begin{algorithmic}[1]
\STATE Specify several $\sigma >0$ values. 
Set the number of bootstrap replicates $B$, say, 10000.
\STATE Compute the adjusted residuals $\hat{\epsilon}$. 
\STATE For each $\sigma$, perform bootstrap resampling to generate $z_\sigma^*$ by $(\ref{eq:resdual})$ for $B$ times and 
compute $\alpha_{\sigma^2}(S | y)=C_S/B$ by counting the frequency $C_S = \#\{j \in \hat{M}_{\sigma}^\ast\text{ and } \hat{s}_{j,\sigma}^\ast=s_j\}$ where 
$\hat{M}_{\sigma}^\ast$ and $\hat{s}_{j,\sigma}^\ast$ are the set of selected features and the estimated sign of feature $j$ by applying the specific algorithm to $(X, z_\sigma^\ast)$, respectively.
Compute $\psi_{\sigma^2}(S|y) = \sigma\bar\Phi^{-1} ( \alpha_{\sigma^2}(S|y))$.
\STATE Estimate parameters $\theta_S(y)$ by fitting model
$\psi_{\sigma^2}(S | y)=\varphi_{S}(\sigma^2 | \theta_S)$.
\STATE Compute the selective $p$-value by
\[
p_{\mathrm{SI}}(H | S, y) = \frac{\bar{\Phi}(z_H )}{\bar{\Phi}( z_H  + z_S  )},
\]
where  $z_H = s_j (a_j^Tz/\|a_j\|_2)/\tau$, $z_S =  \varphi_{S}(0 | \hat{\theta}_S(y))$, and
$\hat{\theta}_S(y)$ is the estimated value of $\theta_S(y)$.
\end{algorithmic}
\end{algorithm}
With the updated computation of $\psi_{\sigma^2}(H|y)$ and $\psi_{\sigma^2}(S|y)$ for regression analysis, we may use Step 3 and Step 4 of Algorithm~\ref{alg:simbp} to 
compute the selective $p$-value.
This becomes Algorithm~\ref{alg:main} for computing an approximately unbiased selective $p$-value for the selected feature $j\in \hat{M}$.
It is worth noting that Algorithm~\ref{alg:main} can be applied to almost any feature selection methods, including MCP and SCAD, in addition to Lasso.
Note also that the multiscale bootstrap is not very sensitive to the choice of the scales.
For $B$, several thousand replications are enough in practice.
The computational cost is the same order as the classical bootstrap method. 
Also note that, in actual implementation of Algorithm~\ref{alg:main}, the Steps 1 to 3 are shared by all the features $j\in \hat M$.
Thus, this algorithm works even for large $p$ such as $p>20$.

Now, we provide the theoretical justification of the proposed algorithm.
Since the boundary surface of the selective region is generally non-smooth as shown in Figure~\ref{fig:lasso}, 
we consider the asymptotic theory of nearly flat surfaces.
In the model space, 
we take the coordinate system $(u,v)\in \mathbb{R}^{n-1} \times \Rb$ such that 
the hypothesis region $H$ can be written by $\{(u,v)\mid v\le 0\}$.
Using this coordinate system, 
let us denote the selective region by $S=\{(u,v)\mid v_s - s(u)\le v\}$ at least locally in a neighborhood of $y$, 
where $v_s\in\mathbb{R}$ and $s$ is a function from $\mathbb{R}^{n-1}$ to $\mathbb{R}$ 
which represents the boundary surface of the selective region.
Here, the $L^1$-norm and $L^\infty$-norm of function $s$ are defined as
$\|s\|_1 = \int |s(u)|\,\mathrm{d}u$ and $\|s\|_\infty = \sup_u|s(u)|$, respectively.
Let $\lambda = \|s\|_\infty$ be the magnitude of the boundary surface $\partial S$ of the selective region.
Even in regression analysis, 
we assume that 
the selection event can be written as $S=\{(u,v)\mid v_s - s(u)\le v\}$ and
the magnitude of the boundary surface $\partial S$ is relatively small at least around the data point $y$ 
in the model space.
That is, we will consider the asymptotic theory in which $\lambda \rightarrow 0$.
In the same way as the local alternative framework, 
this assumption can be interpreted as that
the given surface is on the way to approach the flat surface which is parallel to $\partial H$.
In this paper, this assumption is called nearly flatness of the boundary surface.
In the asymptotic theory of nearly flat surfaces, 
the proposed $p$-value $p_{\mathrm{SI}}(H | S, y)$ has the second-order accuracy.

\begin{theorem}\label{theorem:main}
Let us denote the selective region as $S=\{(u,v)\mid v \ge v_s -s(u) \}$.
Let $\tilde{s}$ be the Fourier transform of the function $s$.
Suppose that the $L^1$-norms $\|s\|_1$ and $\|\tilde{s}\|_1$ are bounded.
Let us assume that $\lambda = \|s\|_\infty$ is sufficiently small.
Then, the selective $p$-value described in Algorithm~\ref{alg:main} has
the second-order accuracy:
\[
\forall \mu \in \partial H;\;\frac{P(p_{\SI}(H|S,Y)<\alpha\mid \mu)}{P(Y\in S\mid \mu)} = \alpha + O(\lambda^{2}).
\]
\end{theorem}
Theorem~\ref{theorem:main} can be considered as a special case of Theorem~5.3 in \citet{TeradaShimodaira17}. 
In the current situation, we can directly obtain the signed distance from $y$ to the boundary surface $\partial H$. 
Thus, the proof of Theorem~\ref{theorem:main} is much simpler than
Theorem~5.3 in \citet{TeradaShimodaira17}. The proof is given in Appendix~\ref{sec:B}.
This theorem provides a theoretical justification for the proposed $p$-value 
when the magnitude of the boundary surface is small, at least locally around the data point.
The assumption requires that the boundary surface $\partial S$ of the selection event is nearly flat and its limiting hyperplane is parallel to the surface $\partial H$ of the hypothesis region, 
at least locally around the data point.
Through the numerical simulations in the next section, 
it seems that this assumption is reasonably satisfied in practice.
Of course, when the two surfaces $\partial H$ and $\partial S$ are not very parallel to each other, 
the proposed $p$-value has a non-negligible  bias.
Solving this problem is an important future work of this research.
%
\begin{remark}
In contrast with the $p$-value $p_\SI(H|S,y)$ in Algorithm~\ref{alg:main},
we also propose a simple selective $p$-value based on the classical bootstrap probability $\alpha_1(S|y)$.
We replace $z_S =  \varphi_{S}(0 | \hat{\theta}_S(y))$ in Step~5 by $z_S' = \bar{\Phi}^{-1}(\alpha_1(S|y))$ to define
\[
p_{\text{SI-BP}}(H|S,y) :=
\frac{\bar{\Phi}(z_H)}{\bar{\Phi}(z_H + z_S')}.
\]
It is also computed with $z_S' = \psi_1(S|y) \approx \varphi_{S}(1 | \hat{\theta}_S(y))$.
If the boundary surface of $S$ is flat, $z_S'=z_S$ and thus $p_{\text{SI-BP}} = p_{\SI}$.
Under the assumption of Theorem~\ref{theorem:main}, 
we can obtain
\[
\forall \mu \in \partial H;\;\frac{P(p_{\text{SI-BP}}(H|S,Y)<\alpha\mid \mu)}{P(Y\in S\mid \mu)} = \alpha + O(\lambda),
\]
indicating that $p_{\text{SI-BP}}$ has a larger bias than $p_{\SI}$.
This result can be proved in much the same way as Theorem~\ref{theorem:main}.
\end{remark}

\section{Numerical experiments}
\label{sec:exp}

Here, we show some numerical experiments to demonstrate the usefulness of our method.
For Lasso, the exact unbiased selective test conditioned on $j\in \hat{M}$ and $\hat{s}_j = s_j\;(j\in \{1,\ldots,p\},\, s_j\in \{+, -\})$ can be constructed (\citealp{LeeEtAl16, LiuEtAl18}).
At first, we will show that our selective $p$-value with multiscale bootstrap (i.e., $p_\SI$) approximates the exact selective $p$-value for Lasso.
Here, Lasso is defined as $\min_{\beta\in \Rb^p} \|z-X\beta\|_2^2/2 + \sum_{j=1}^p\rho |\beta_j|$.
Set $(n,p) = (50,25)$ and 
$\beta=(2,2,2,2,2,0,\dots,0)^T\in \Rb^p$; $\beta_j=0$ for the features $j=6,\ldots,25$.
The elements $x_{ij}$ of the input matrix $X$ were independently generated from the standard normal distribution $N(0,1)$.
Then, the response $z\in \Rb^n$ was generated as $z = X\beta + \epsilon$, where $\epsilon$ was generated from the $n$-dimensional standard normal distribution $N_n(0_n,I_n)$.
In our algorithm, we used $\sigma^2= 0.5,0.6,\dots,1.5$ as the scales and $B=10^4$ as the number of bootstrap replicates.
Here, we note that the choice of scales has only little effect on the stability of the result.
The experiment about this point can be found in Appendix~\ref{app:C}.
We simulated $2000$ independent datasets.
We set the significance level $\alpha = 5\%$. 
For computing false positive rates accurately, 
the variables with zero coefficients need to be selected several hundred times. 
Here, we used Lasso with the penalty parameter $\rho = 10$ as the feature selection method. Each variable with zero coefficient was selected approximately $250$ times out of $2000$ datasets in this experiment. 
In each dataset, we performed the classical $t$-test, the selective (one-sided) test conditioned on $\hat{M} = M$ and $\hat{s}_M=s_M$ for $M\subset\{1,\ldots,p\},\, s_M\in\{+,-\}^{|M|}$ (\citealp{LeeEtAl16}; only for Lasso),  the selective (two-sided) test conditioned on $j\in\hat{M}$ for $j\in \{1,\ldots,p\}$ (\citealp{LiuEtAl18}; only for Lasso) and our approximately unbiased (one-sided) test with multiscale bootstrap for each selected feature.
We count how many times, say $N_j$, the feature $j$ is selected. 
For each test, we also count how many times (say $R_j$)
the null hypothesis $H_0:\beta_j  \lesseqgtr \eta$ is rejected, 
and the selective rejection probability is estimated by $R_j/N_j$.

\begin{figure}[h!]
\centerline{ 
 \begin{minipage}{0.48\textwidth}
	 \centering
  	\includegraphics[width=.98\textwidth]{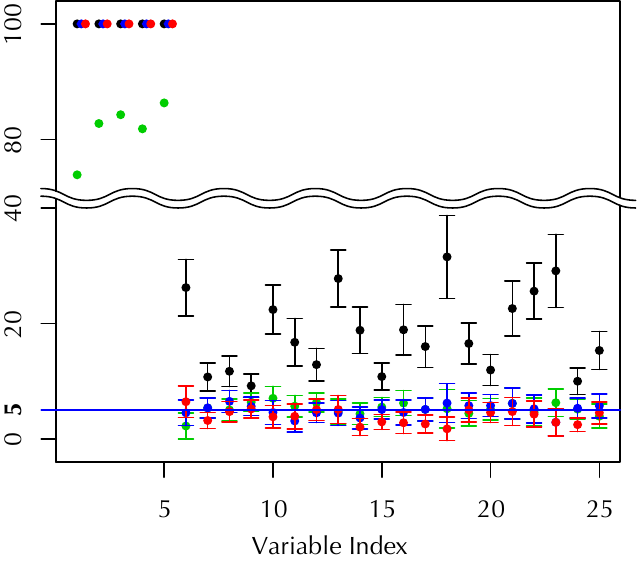}
	(a)~Lasso ($n= 50,\; p = 25$)
  \end{minipage}
 \begin{minipage}{0.5\textwidth}
	\centering
 	 \includegraphics[width=\textwidth]{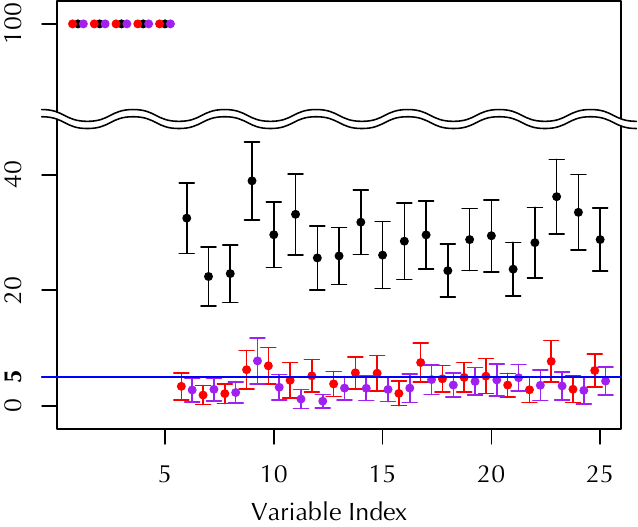}
	(b)~MCP ($n= 50,\; p = 25$)
  \end{minipage}
 }
 \vspace{10pt}
 \centerline{ 
   \begin{minipage}{0.49\textwidth}
	\centering
 	 \includegraphics[width=.98\textwidth]{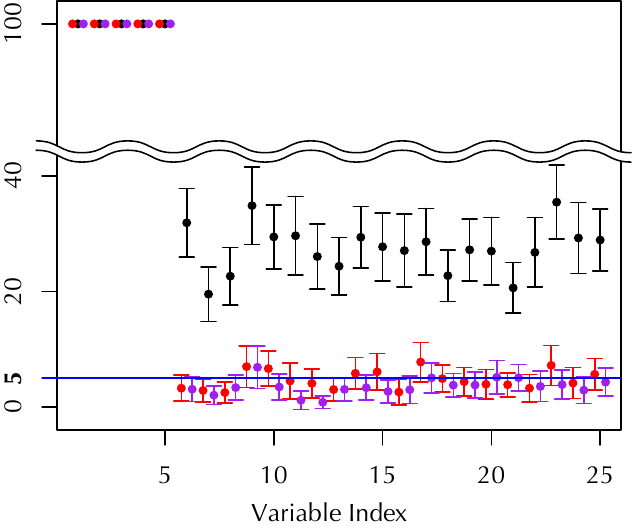}
	(c)~SCAD ($n= 50,\; p = 25$)
  \end{minipage}
     \begin{minipage}{0.49\textwidth}
	\centering
 	 \includegraphics[width=.98\textwidth]{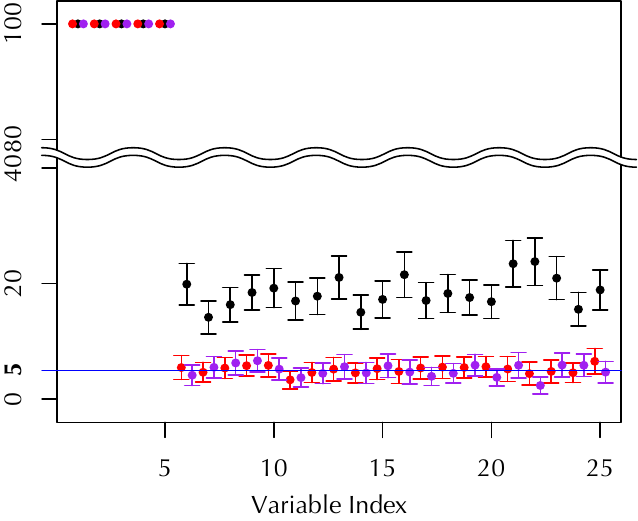}
	(d)~MCP ($n= 100,\; p = 25$)
  \end{minipage}
  }
  \caption{Selective rejection probabilities (in percent) with two standard deviations for each feature.
  Black: the classical $t$-test, 
  Green: the exact selective test conditioned on $\hat{M} = M$ and $\hat{s}_M=s_M$ (\citealp{LeeEtAl16}; only for Lasso),
  Blue: the exact selective test conditioned on $\{j \in \hat{M}\}$ (\citealp{LiuEtAl18}; only for Lasso),
  Red: our approximately unbiased test with the multiscale bootstrap conditioned on $j\in \hat{M}$ and $\hat{s}_j = s_j$ (i.e., $p_\SI$), Purple (for MCP and SCAD): our approximately unbiased test with the classical bootstrap conditioned on $j\in \hat{M}$ and $\hat{s}_j = s_j$ (i.e., $p_{\text{SI-BP}}$).
  }
  \label{fig:exp1}
\end{figure}

The panel (a) of Figure~\ref{fig:exp1} shows the selective rejection probabilities of each feature for Lasso, where the four test methods are compared.
In this plot, we can see that the selective rejection probabilities of our test with $p_\SI$ for the features $6$ to $25$ with $\beta_j=0$ are around $5\%$. 
Thus, it is shown that our multiscale bootstrap method approximately satisfies the unbiasedness in the sense of the equation~$(\ref{eq:unbiased})$.
We can also see that the classical inference does not provide a valid inference after feature selection; the classical $t$-test gives more false positives than expected from the specified $\alpha$ level.
Moreover, instead of the Lasso penalty $\rho|\beta_j|$, 
we also used the following MCP and SCAD penalties with the tuning parameter $(\rho,\gamma)=(10,3.7)$ as the feature selection methods:
\ba
\mathrm{MCP}(\beta_j|\rho, \gamma)
&=
\begin{cases}
\rho|\beta_j| - \beta_j^2/(2\gamma) & (|\beta_j|\le \gamma\rho)\\
\gamma \rho^2/2 & (|\beta_j|>\gamma \rho)
\end{cases},\\
\mathrm{SCAD}(\beta_j|\rho, \gamma)
&=
\begin{cases}
\rho|\beta_j| & (|\beta_j|\le \rho)\\
\frac{2\gamma \rho|\beta_j|-\beta_j^2-\rho^2}{2(\gamma - 1)} & (\rho <|\beta_j|<\gamma \rho)\\
\rho^2(\gamma+1)/2 & (|\beta_j|\ge \gamma\rho)
\end{cases}.
\ea

For non-convex penalties, we have local minimum and multiple global minimum issues.
In general, we assume that the selection event can be represented as the fixed set in the data space. Thus, both issues have unexpected effects in selective inference. 
To overcome these issues, we used the fixed initial values for MCP and SCAD
\footnote{Another approach is extending the data space to include the space of initial values.}.
In this experiment, we used the R package \texttt{plus} (\citealp{plus}) for MCP and SCAD.
The algorithm of this package generates a piecewise linear path of coefficients, starting with zero coefficients for infinity penalty.

Our multiscale bootstrap method works well not only for Lasso but also for more complicated feature selection methods such as MCP and SCAD.
The panels (b) and (c) of Figure~\ref{fig:exp1} show the selective rejection probabilities in the cases of MCP and SCAD, respectively.
Here, we note that the existing selective inference methods (Green, Blue) cannot be applied to MCP and SCAD.
Each feature with zero coefficient was selected approximately $250$ times by MCP and SCAD in this experiment.
In this setting, whereas no exact unbiased selective inference is proposed, 
the selective rejection probabilities of our test for the features $6$ to $25$ are around $5\%$.
In general, we can get a more accurate result as the sample size increases. 
In fact, the panel (d) of Figure~\ref{fig:exp1} is the result of the same experiment about MCP with larger sample size $n = 100$.
Compared with the case of $n = 50$ of the panel (b),
we can see the more accurate result (i.e., less variations) in the case of $n = 100$.

Moreover, we also computed the simpler selective $p$-value $p_{\text{SI-BP}}$ based on the classical bootstrap in both settings with MCP and SCAD.
For MCP with $n=50$, 
the selective rejection probabilities of $p_{\text{SI}}$ and $p_{\text{SI-BP}}$ 
under the null hypotheses are $4.62\%$ and $3.45\%$, respectively.
For SCAD with $n=50$, 
the selective rejection probabilities of $p_{\text{SI}}$ and $p_{\text{SI-BP}}$ 
under the null hypotheses are $4.62\%$ and $3.52\%$, respectively.
These results show that the multiscale bootstrap method is more accurate than the classical bootstrap method in accordance with theory.
In the setting with $n=100$, 
the selective rejection probabilities of $p_{\text{SI}}$ and $p_{\text{SI-BP}}$ 
under the null hypotheses are $5.16\%$ and $5.02\%$, respectively.
As the sample size increases, 
both $p_{\text{SI}}$ and $p_{\text{SI-BP}}$ provide almost unbiased results.
It could be because the boundary surface of the selection event is almost flat, at least locally around the data point, in a larger sample case.

\begin{figure}[h]
 \centering
  \includegraphics[keepaspectratio,width=.55\textwidth]{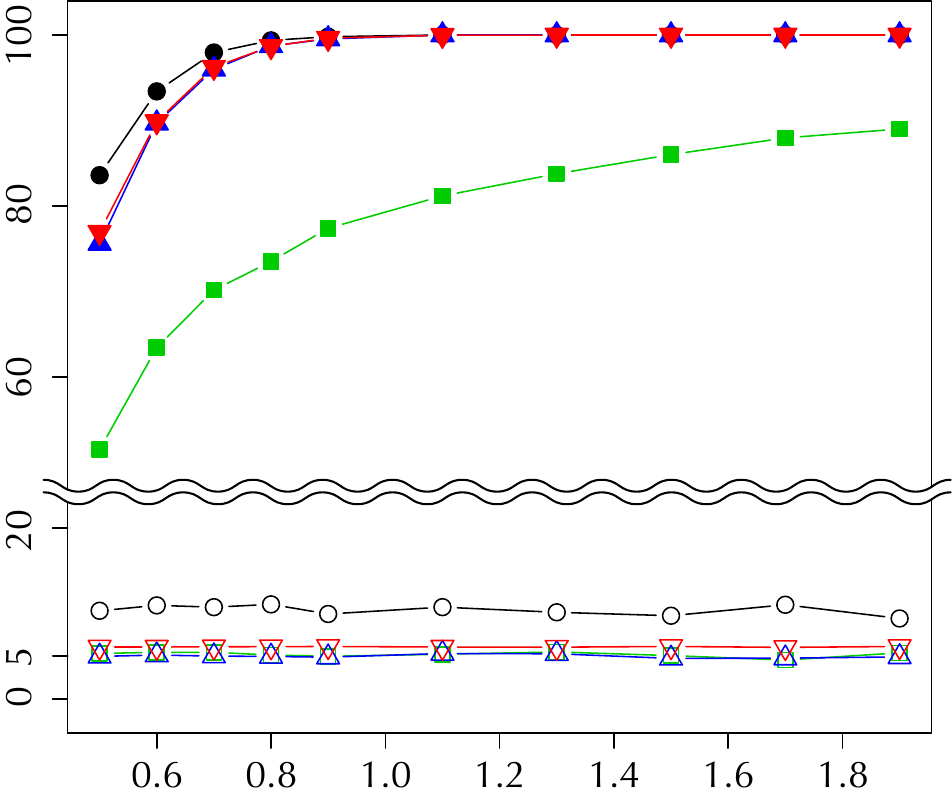}
  \caption{True positive rates (upper) and false positive rates (lower) of four tests. Colors correspond to the tests as shown in the panel (a) of Figure~\ref{fig:exp1}. 
  The horizontal axis represents the strength $\theta$ of the weights for the non-zero features. 
  }
  \label{fig:power}
\end{figure}

In addition, set 
$\beta=(\theta,\theta,\theta,\theta,\theta,0,\dots,0)^T\in \Rb^p$ and $(n,p) = (50,10)$.
We compare the true positive rates (TPRs; i.e., statistical powers) and the false-positive rates (FPRs; i.e., type-I errors) of these tests with changes of $\theta$ 
in the case of Lasso. 
Here, TPR is defined by the proportion of selected non-zero features that are correctly identified, 
and FPR is defined by the proportion of selected zero-features that are incorrectly detected.
Figure~\ref{fig:power} shows that 
both the proposed method via multiscale bootstrap and the exact selective test conditioned on $j\in\hat{M}$ \citep{LiuEtAl18} 
not only have desirable high TPRs but also control FPRs at the significance level $\alpha=5\%$.
For the non-selective $t$-test, the FPR is not controlled, whereas the highest TPR is attained.
The $t$-test (the black line) does not control the false positive rate. Thus it is not valid.
Focusing on the TPR of the \emph{over-conditioning} selective test \citep{LeeEtAl16}, 
we can see that the unnecessarily restrictive selection event $\{\hat{M} = M,\hat{s}_M=s_M\}$ leads to the lower statistical power.

Next, we deal with the prostate cancer data (\citealp{StameyEtAl89}), which is available in the R package \texttt{ElemStatLearn} (\citealp{ElemStatLearn2015}).
\citet{StameyEtAl89} studied the relation between the level of prostate-specific antigen (PSA) and $8$ clinical measures: 
the log cancer volume (\texttt{lcavol}), the log prostate weight (\texttt{lweight}), and so on.
Here, we consider a linear regression model to the log of PSA (\texttt{lpsa}) with $8$ clinical measures. In this application, we prepossessed the data so that each feature has a mean zero and unit variance.
\begin{figure}[h]
 \centering
  \includegraphics[keepaspectratio,width=.55\textwidth]{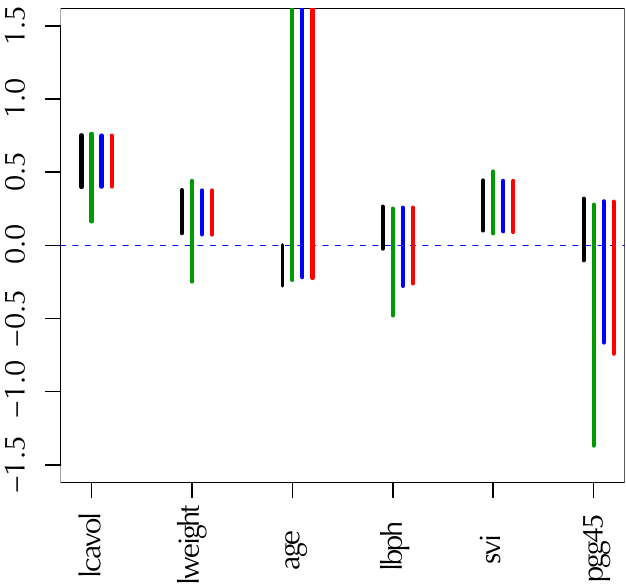}
  \caption{Various confidence intervals for the coefficients of the selected features. 
  (Black) the non-selective CIs $[L_j^{(a)}, U_{j}^{(a)}]$. 
  (Green) the selective CIs $[L_j^{(b)}, U_{j}^{(b)}]$. 
  (Blue) the selective CIs $[L_j^{(c)}, U_{j}^{(c)}]$.
  (Red) our approximate selective CIs $[L_j^{(d)}, U_{j}^{(d)}]$ with the multiscale bootstrap.}
  \label{fig:CI}
\end{figure}
The main purpose is to provide the selective confidence intervals (CIs) for the coefficients of the $6$ selected features by Lasso with the penalty $\rho = 5$.
Here, we also set $\alpha = 5\%$.
We computed four types of confidence intervals with confidence level $1-\alpha$ as shown in Figure~\ref{fig:CI}:
the non-selective CI $[L_j^{(a)}, U_{j}^{(a)}]$ using $t$-distribution, the selective CI $[L_j^{(b)}, U_{j}^{(b)}]$ conditioned on $\{\hat{M}=M, \hat{s}_M=s_M\}$ (\citealp{LeeEtAl16}), 
the selective CI $[L_j^{(c)}, U_{j}^{(c)}]$ conditioned on $\{j\in\hat{M},\hat{s}_j=s_j\}$ (\citealp{LiuEtAl18}), and 
the approximate selective CI $[L_j^{(d)}, U_{j}^{(d)}]$ based on our approximately unbiased $p$-values $p_\SI$ based on the multiscale bootstrap.
We note that the first three CIs satisfy the following equations, respectively, for $j\in\{1,\ldots,p\}$, $s_j \in\{+,-\}$, 
$M\subseteq \{1,\dots,p\}$, $s_M\in \{+,-\}^{|M|}$:
\ba
&P(\beta_j \in [L_j^{(a)}, U_{j}^{(a)}]) = 1-\alpha,\; P(\beta_j \in [L_j^{(b)}, U_{j}^{(b)}]\mid \hat{M} = M, \hat{s}_M=s_M) = 1-\alpha,\\
&P(\beta_j \in [L_j^{(c)}, U_{j}^{(c)}] \mid j\in \hat{M}, \hat{s}_j=s_j) = 1-\alpha.
\ea
From the plot, we can see that 
our selective CIs $[L_j^{(d)}, U_{j}^{(d)}]$ approximates the exact selective CIs $[L_j^{(c)}, U_{j}^{(c)}]$ very well. 
Moreover, the over-conditioning of the  selection event $\{\hat{M}=M,\hat{s}_M=s_M\}$ made CIs $[L_j^{(b)}, U_{j}^{(b)}]$ wider than $[L_j^{(c)}, U_{j}^{(c)}]$, and 
this indicates that the less restrictive selection event $\{j\in \hat{M}, \hat{s}_j=s_j\}$ is preferable.

\section{Discussion}

A new multiscale bootstrap method (Algorithm~\ref{alg:main}) is proposed to compute approximately unbiased selective $p$-values and confidence intervals for regression coefficients after feature selection. The new method is useful in particular for complicated feature selection algorithms such as MCP and SCAD, while existing methods are only available for simpler feature selection algorithms such as Lasso.
The new method also computes shorter confidence intervals than most existing methods by minimally-conditioning on each selected feature instead of over-conditioning on all selected features.

\begin{figure}[h]
 \centering
  \includegraphics[keepaspectratio,width=.55\textwidth]{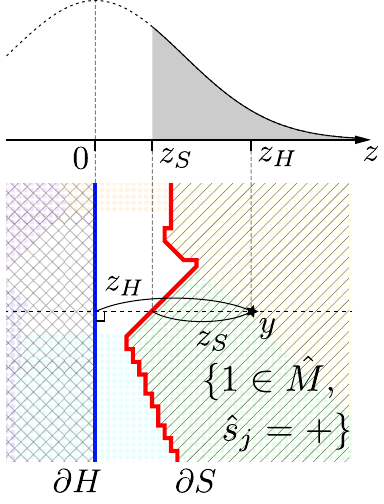}
  \caption{The exact selective inference for the hypothesis region with the flat boundary.}
  \label{fig:geometric_rep}
\end{figure}

The proposed method is closely related to the exact selective inference such as \cite{LeeEtAl16} and \cite{LiuEtAl18}.
Here, in addition to the Gaussian assumption, we assume that the boundary surface of the hypothesis region is flat. 
Let us consider the line passing through the point $y$ and perpendicular to the boundary of $H$. 
By setting the projection point $\hat{\mu}(y) \in \partial H$ of $y$ as the origin, we can consider the one-dimensional coordinate system $z$ on the line. If we know
the distance from $y$ to $\partial H$ as well as intervals representing the intersection of the line and the selective region $S$, we can perform the exact selective inference. For example, in Figure~\ref{fig:geometric_rep}, 
the distance is $z_H$ and
the interval is $[z_S,\infty)$.
As with the polyhedral lemma, the following $p$-value provides the exact selective inference:
\[
p(y) = \frac{\bar{\Phi}(z_H)}{\bar{\Phi}(z_H+z_S)}.
\]
The explicit forms of the intervals can be obtained for the Lasso case, but it may not be possible for more complicated cases.
In the proposed method, we estimate the geometric quantities $z_H$ and $z_S$ indirectly via the multiscale bootstrap.
Alternating to this approach, we can use the grid search on the one-dimensional coordinate system $z$ to obtain the intervals.
In association with this approach,
\cite{DuyTakeuchi20} proposes a parametric programming-based method that can perform the exact selective inference for the Lasso very efficiently without conditioning on signs. 

Our method is applicable to the case of $p>n$, provided that $(X^TX)^+X^T\mu$ is the target of inference, where $A^+$ denotes the pseudo-inverse matrix of $A$.
However, in this case, 
it is difficult to estimate the error variance tau or the residuals reasonably well.
Thus, the selective inference in the case of $p > n$ is important future work.
In theory, we may consider the application of the proposed framework for the submodel setting.
However, 
the selection event $\{\hat{M}=M\}$ is often too small, i.e., the bootstrap probability becomes very small for the selection event, 
and thus 
the proposed framework does not work well.
Combining with the randomized response method by \cite{TianTaylor17+},
we may propose an appropriate selective inference based on the multiscale bootstrap method for the submodel setting.

In this paper, the multiscale bootstrap is used only for the selective region, because the hypothesis region with a flat boundary surface (i.e., $\beta_j=0$) is easily expressed in the model space. However, Algorithm 1 is valid even for a general hypothesis region with a curved surface. Therefore, we may extend our method for, say, non-linear regression or multiple comparisons of regression coefficients in future work.


\section*{Acknowledgments}
The authors would like to thank an associate editor and two reviewers for valuable comments and suggestions that improve the quality of the paper considerably.
This research was supported in part by JSPS KAKENHI Grant (JP16K16024, JP20K19756, and JP20H00601 to YT; JP16H02789, JP20H04148, and JP20H04243 to HS) and 
MEXT Project for Seismology toward Research Innovation with Data of Earthquake (STAR-E) Grant Number JPJ010217 (to YT).

\bibliographystyle{imsart-nameyear}     
\bibliography{simbp_bib,stat2017}

\newpage
\appendix
\section{Asymptotic theory for nearly flat surfaces}
\label{sec:A}

In Section~\ref{sec:mbp}, we only describe the multiscale bootstrap method 
for the hypothesis and selective regions with smooth boundaries.
In the classical large sample theory, an important point is that 
the smooth boundary surface of the hypothesis region approaches a flat surface in a neighborhood of any point on its boundary surface.
However, this claim cannot be true for regions with non-smooth boundaries since cone-shaped regions are scale-invariant in the neighborhood of the vertex.
In many practical situations, the hypothesis and selective regions could have non-smooth surfaces.
Thus, \citet{Shimodaira08} develops a new theoretical framework, 
called the asymptotic theory of {\it nearly flat surfaces}.
In this theory, we consider the situation that the magnitude of boundary surfaces, say $\lambda$, becomes small, 
that is, any boundary surfaces approach flat surfaces at least locally in a neighborhood.
The artificial parameter $\lambda$ is introduced, and consider the situation of $\lambda \rightarrow 0$ instead of $n\rightarrow \infty$.
More precisely, suppose that the $L^1$-norms of function $h$ and its Fourier transform $\tilde{h}$, i.e.,  $\|h\|_1$ and $\|\tilde{h}\|_1$, are bounded and that the $L^\infty$-norm $\|h\|_\infty$ of $h$ has the same order as $\lambda$.
Here, the function satisfying these properties is called nearly flat.
Then, we consider the asymptotic theory as $\lambda \rightarrow 0$.
Note that $\lambda$ in this theory is corresponding to $1/\sqrt{n}$ in the classical large sample theory.
Here, we assume that
the hypothesis and selective regions are defined as follows, respectively:
\ba
H &= \{(u,v)\mid v \le -h(u), u\in \Rb^m, v\in \Rb\},\text{ and }\\
S &= \{(u,v)\mid v_s-s(u) \le v, u\in \Rb^m, v\in \Rb\},
\ea
where $h$ and $s$ are nearly flat functions, and $v_s\in\Rb$.

Even in this theory, the bootstrap probability also has the first-order accuracy:
\[
\forall \mu \in \partial H;\;P(p_{\BP}(H|Y)<\alpha\mid \mu) = \alpha + O(\lambda).
\]
Write $y=(u,v)\in \Rb^{m}\times \Rb$. Then, the distribution of the bootstrap sample $Y^\ast=(U^\ast,V^\ast)$ with scale $\sigma^2$ is given as
\[
U^\ast \sim N_m(u,\sigma^2 I_m),\quad V^\ast \sim N(v,\sigma^2).
\]
Let $\Ecal_{\sigma^2}$ denote the expectation operator related to $U^\ast$, that is,
\[
\Ecal_{\sigma^2} h(u) = E_{\sigma^2}[ h(U^\ast)| u] = \Fcal^{-1}[ e^{-\sigma^2\|\omega\|^2/2}\tilde{h}(\omega) ](u),
\]
where $E_{\sigma^2}[\cdot|u ]$ is the expectation related to $U^\ast$ and $\Fcal^{-1}$ is the inverse Fourier transform operator.
For the normalized bootstrap $z$-value, we have the following scaling-law which is parallel to one of the large sample theory:
\[
\psi_{\sigma^2}(H|y) = v + \Ecal_{\sigma^2} h(u) + O(\lambda^{2}).
\]
We note that, for $\sigma_1^2,\sigma_2^2>0$, it follows that $\Ecal_{\sigma_1^2}\Ecal_{\sigma_2^2}h(u) = \Ecal_{\sigma_1^2+\sigma_2^2}h(u)$.
Hence, at least formally, the expected value with a negative variance is defined as
\[
\Ecal_{\sigma^2}^{-1} h(u) = \Ecal_{-\sigma^2} h(u)= \Fcal^{-1}[ e^{\sigma^2\|\omega\|^2/2}\tilde{h}(\omega) ](u).
\]
Note that $\Ecal_{-\sigma^2} h(u)$ may not be well-defined in general.
For a detailed discussion about $\Ecal_{-\sigma^2} h(u)$, we refer the reader to \citet{Shimodaira08}.
If $\Ecal_{-1} h(u)$ can be defined, the $p$-value $p_{\AU}(H|y) =
\bar{\Phi}(\psi_{-1}(H|y)) = \bar{\Phi}(v + \Ecal_{-1}h(u))$ has 
the second-order accuracy for non-selective test (\citealp{Shimodaira08}):
\[
\forall \mu \in \partial H;\;P(p_{\AU}(H|Y)<\alpha\mid \mu) = \alpha + O(\lambda^{2}).
\]
As with the classical large sample theory, 
if $\Ecal_{-1} h^2(u)$ also exits,
it can be shown that 
$p_{\AU}(H|y)$ has the third-order accuracy 
with bias only $O(\lambda^{3})$.

For the smooth $h$, it follows that $\Ecal_{\sigma^2} h(u)=\sum_{j=0}^\infty \sigma^{2j}\theta_j(u)$.
That is, letting $\theta_{H,0} = v + \theta_0(u)$ and $\theta_{H,j} = \theta_j(u)\;(j\ge 1)$, 
$\varphi_H(\sigma^2|\theta_H)$ can be modeled as $\theta_{H,0} + \theta_{H,1}\sigma^2 + \theta_{H,2}(\sigma^2)^2 + \cdots$.
Thus, using a polynomial regression with predictor $\sigma^2$, we can compute the $p$-value $p_{\AU}(H|y) = \bar\Phi(\theta_{H,0} - \theta_{H,1} + \theta_{H,2} - \cdots)$ by formally letting $\sigma^2 = -1$.
In contrast, for a cone-shaped region $H$, it is shown that $\Ecal_{\sigma^2} h(u)=\sum_{j=0}^\infty \sigma^{1-j}\theta_j(u)$.
Since we have $\beta_j(u) = O(\|u\|^j)$ as $\|u\|\rightarrow 0$, 
focusing on first two terms, we obtain
\[
\Ecal_{\sigma^2} h(u) \approx \theta_0(u) \sqrt{\sigma^2} + \theta_1(u).
\]
In this model, we cannot take $\sigma^2=-1$, and $\Ecal_{-1} h(u)$ does not exist for a cone-shaped region $H$.
This observation is related to the important fact proved by \citet{Lehmann:1952:TMH} that an unbiased test cannot exist for a cone-shaped hypothesis region.
Set $\theta_{H,0} = v + \theta_1(u)$ and $\theta_{H,1}=\theta_0(u)$, 
and then the normalized bootstrap $z$-value $\psi_{\sigma^2}(H|y)$ can be approximated by the model $\theta_{H,0}+\theta_{H,1}\sigma$; note the predictor is $\sigma=\sqrt{\sigma^2}$ instead of $\sigma^2$.
Here, we also denote by $\varphi_H(\sigma^2|\theta_H)$ the model which approximates the normalized bootstrap $z$-value $\psi_{\sigma^2}(H|y)$.
For fixed $\sigma_0^2>0$, let $\varphi_{H,k}(\sigma^2|\theta_H,\sigma_0^2)$ be 
the truncated Taylor expansion of $\varphi_H(\sigma^2|\theta_H)$ with the first $k$ terms at $\sigma_0^2$:
\banum
\varphi_{H,k}(\sigma^2|\theta_H,\sigma_0^2)
=
\sum_{j=0}^{k-1}\frac{(\sigma^2 - \sigma_0^2)^j}{j!}
\frac{\partial^j \varphi_H(\sigma^2|\theta_H)}{\partial (\sigma^2)^j}\biggm|_{\sigma^2=\sigma_0^2}.
\label{eq:A.1}
\eanum
We can always use the above formula for extrapolating $\psi_{\sigma^2}(H|y)$ to $\sigma^2 \le 0$.
Therefore, Algorithm~~\ref{alg:simbp} is updated to Algorithm~\ref{alg:simbp2}.
In practice, Algorithm~~\ref{alg:simbp2} with $k=2$ can be simply implemented as  Algorithm~~\ref{alg:simbp} with the linear model $\theta_{H,0}+\theta_{H,1}\sigma^2$ and a narrow range of $\sigma^2=n/n'$ values around $\sigma^2_0$.

\begin{algorithm}[h]
\caption{Computing approximately unbiased $p$-values for general regions $H$ and $S$ (applicable to non-smooth boundary surfaces)}
\label{alg:simbp2}
\begin{algorithmic}[1]
\STATE Same as Algorithm~\ref{alg:simbp}
\STATE Same as Algorithm~\ref{alg:simbp}
\STATE Same as Algorithm~\ref{alg:simbp}
\STATE Approximately unbiased $p$-values of non-selective inference ($p_\mathrm{AU}$) and of
selective inference ($p_\mathrm{SI}$) are computed as follows.
Specify $k\in \Nb$, $\sigma_0^2,\sigma_{-1}^2>0$ (e.g., $k=3$ and $\sigma_{-1}^2=\sigma_0^2=1$).
Compute $p$-values by $p_{\mathrm{AU},k}(H|y) = \bar{\Phi}(z_{H,k})$ and 
$p_{\mathrm{SI},k}(H | S, y) = \bar{\Phi}(z_{H,k})/\bar{\Phi}( z_{H,k}+ z_{S,k})$,
where $z_{H,k} = \varphi_{H,k}(-1 | \hat{\theta}_H(y),\sigma_{-1}^2)$ and
$z_{S,k} =  \varphi_{S,k}(0 | \hat{\theta}_S(y),\sigma_0^2)$ computed by formula (\ref{eq:A.1}).
\end{algorithmic}
\end{algorithm}

For fixed $k \in \Nb$, we consider the following $p$-value:
\[
p_{\AU,k}(H|y) = \bar{\Phi}(\varphi_{H,k}(-1|\theta_H,\sigma_0^2)).
\]
Under some regularity conditions, \citet{Shimodaira08} proves that
\[
\lim_{k\to\infty} P(p_{\AU,k}(H|Y)<\alpha\mid \mu) = \alpha + O(\lambda^{2})
\]
at each $\mu \in \partial H$.
Moreover, for general selective inference with possibly non-smooth boundary surfaces,
\citet{TeradaShimodaira17} proposes the following selective $p$-value:
\[
p_{\SI,k}(H|S,y) = \frac{\bar{\Phi}(\varphi_{H,k}(-1|\theta_{H},\sigma_{-1}^2))}{\bar{\Phi}(\varphi_{H,k}(-1|\theta_{H},\sigma_{-1}^2) + \varphi_{S,k}(0|\theta_{S},\sigma_0^2))},
\]
where $\sigma_{-1}^2,\sigma_0^2>0$.
In addition, it is shown that the selective $p$-value has the second-order accuracy:
\[
\lim_{k\to\infty} \frac{P(p_{\SI,k}(H|S,Y)<\alpha\mid \mu)}{P(Y\in S\mid \mu)} = \alpha + O(\lambda^{2})
\]
at each $\mu \in \partial H$.

\section{Proof of Theorem~\ref{theorem:main}}
\label{sec:B}

\begin{proof}
First, we show that, for given $\alpha \in (0,1)$, 
there exists a nearly flat function $r$ such that
\banum
\forall \mu \in \partial H;\;\frac{P(Y\in R\mid\mu)}{P(Y\in S\mid\mu)} = \alpha + O(\lambda^2),
\label{eq:unbiasedness}
\eanum
where $R=\{(u,v)\mid v > v_r - r(u)\}$ and $v_r = \bar{\Phi}^{-1}(\alpha\bar{\Phi}(v_s))$.
By Lemma 5.1 in \citet{TeradaShimodaira17} or equivalently eq.~(5.3) in \citet{Shimodaira08}, we have
\ba
P(Y \in S\mid\mu) 
&= 1-\Phi(v_s - \mathcal{E}_1s(\theta) )+ O(\lambda^2)\\
&= 1- \{\Phi(v_s)-\phi(v_s)\mathcal{E}_1s(\theta)\}+O(\lambda^2)
\ea
for $\mu=(\theta,0) \in \partial H$.
Let us temporarily assume that $r$ is nearly flat. 
Then, we also have
\[
P(Y \in R\mid\mu) = 1- \{\Phi(v_r) - \phi(v_r)\mathcal{E}_1r(\theta)\}+ O(\lambda^2).
\]
From Eq.~(\ref{eq:unbiasedness}), it follows that 
\ba
\bar{\Phi}(v_r) + \phi(v_r)\Ecal_1 r(\theta) + O(\lambda^2)
= \alpha\left[\bar{\Phi}(v_s) + \phi(v_s)\mathcal{E}_1s(\theta) + O(\lambda^2)\right].
\ea
Since $\bar{\Phi}(v_r)= \alpha\bar{\Phi}(v_s)$,
we have
$\Ecal_1 r(\theta)=\alpha C\mathcal{E}_1s(\theta) + O(\lambda^2)$, 
where $C=\phi(v_s)/\phi(v_r)$.
Thus, applying the inverse operator $\Ecal_{-1}$ to both sides, 
we obtain $r(u)=\alpha C s(u) + O(\lambda^2)$.
Since $s$ is nearly flat, $r$ should be nearly flat. 
Similarly, we can show that the above $r$ actually satisfies Eq.~(\ref{eq:unbiasedness}).
Combining $\phi(v_r)r(u) =  \alpha \phi(v_s)s(u) + O(\lambda^2)$ with $\bar{\Phi}(v_r)= \alpha\bar{\Phi}(v_s)$,
we obtain $\bar{\Phi}(v_r) + \phi(v_r)r(u) = \alpha\left\{ \bar{\Phi}(v_s) + \phi(v_s)s(u)\right\} + O(\lambda^2)$.
By Taylor's theorem, 
we deduce that
\banum
\bar{\Phi}(v_r - r(u)) = \alpha\bar{\Phi}(v_s - s(u)) + O(\lambda^2).
\label{eq:alpha}
\eanum

Now, we consider the rejection region $R$ based on $p_{\mathrm{SI}}$, that is, 
$R=\{(u,v)\mid p_{\mathrm{SI}}(H|S,y)<\alpha\}$.
By Lemma 5.1 in \citet{TeradaShimodaira17} or equivalently eq.~(5.3) in \citet{Shimodaira08}, for $y=(u,v)$, 
we have $\psi_{\sigma^2}(S|y)= -v + v_s - \Ecal_{\sigma^2}s(u) + O(\lambda^2)$.
Since $\Ecal_{0}s(u) = s(u)$, we have $z_S = \psi_{0}(S|y)= -v + v_s - s(u) + O(\lambda^2)$.
Let us recall that $\psi_{\sigma^2}(H|y) = v(H|y) = v$ for $H=\{(u',v') \mid  u'\in\mathbb{R}^m, v' \le 0\}$, and so $z_H = v.$
Thus, we obtain
\[
p_{\SI}(H|S,y) =
\frac{\bar\Phi(z_H)}{\bar\Phi(z_H + z_S)} = 
\frac{\bar{\Phi}(v)}{\bar{\Phi}(v_s - s(u))} + O(\lambda^2).
\]
By substituting $v = v_r-r(u)$ above, it follows from Eq.~(\ref{eq:alpha}) that $p_{\SI}(H|S,y) = \alpha + O(\lambda^2)$ for $y=(u,v_r-r(u)) \in \partial R$.
This finishes the proof.
\end{proof}

\section{Choice of the tuning parameters in multiscale bootstrap}\label{app:C}

\begin{figure}[h]
\begin{center}
	\includegraphics[keepaspectratio,width=0.5\textwidth]{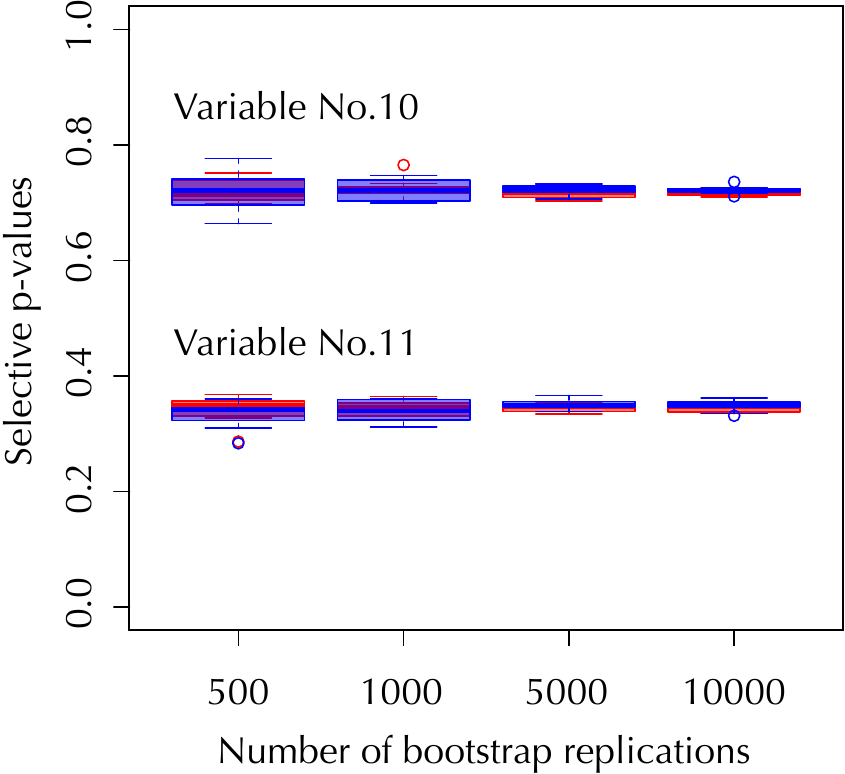}
\end{center}
\caption{Stability of multiscale bootstrap with the choice of the scales and the number of replications.}
\label{fig:sta}
\end{figure}

The multiscale bootstrap is not very sensitive to the choice of the scales. 
For confirming this fact, we additionally performed experiments with two settings of scales, as shown in Figure~\ref{fig:sta}. 
We choose $\ell$ ($=5$ and 10) scales from the interval between 0.1 and 2 with equal spaces in the log-scale. 
We also changed the number of replications $B$ ($= 500, 1000, 5000, 10000$) in multiscale bootstrap. 
The other parameters are the same as the experiment in Section~\ref{sec:exp}.
For a simulated data, we computed selective $p$-values under the various settings ($B = 500, 1000, 5000, 10000; \ell = 5, 10$) of multiscale bootstrap. 
Under each setting, we computed the selective $p$-value 10 times for two selected features No. 10 and No. 11 whose true coefficients are zero. 

In Figure~\ref{fig:sta}, the red box plots correspond to the setting $\ell = 5$ and the blue ones to the setting $\ell = 10$.
The averages of 10 $p$-values are almost the same among all settings, and thus the $p$-values are not sensitive to $\ell$. 
On the other hand, the variance of the $p$-values decreases as $B$ increases. 
Several thousand replications are enough in practice. 

\end{document}